# Security Analysis of Two Recent Pairing-Free Certificateless Two-Party Authenticated Key Agreement Protocols for Smart Grid

**Yong-Jin Kim**[1], **Dok-Jun An**[1], **Son-Gyong Kim**[2], **Kum-Sok Sin**[3], **You-Jin Jong**[4], and **Ok-Chol Ri**[4]

[1]Faculty of Mathematics, KIM IL SUNG University, Pyongyang, 999093, D.P.R of Korea
[2]Institute of Management Practice, Ministry of Information Industry, Pyongyang, 99903, D. P. R of Korea
[3]Pyongyang Software Joint Development Center, Pyongyang, 999093, D.P.R of Korea
[4]Kum Sung Middle School Number 2, Pyongyang, 999093, D.P.R of Korea

Corresponding author: Yong-Jin Kim (kyj0916@126.com)

**ABSTRACT:** Smart grids are intelligent power transmission networks that monitor and control communication participants and grid nodes to ensure bidirectional flow of information and power between all nodes. To secure the smart grid, it is very important to design the key agreement protocol. The pairing-free certificateless two-party authenticated key agreement protocol has been widely studied and applied as a basic core protocol to protect the security of the smart grid. Until now, various protocols have been proposed, and these protocols are being introduced and operated not only in smart grid, but also in smart cities, healthcare, and vehicle ad hoc networks. In this paper, we analyzed the security properties of two recently proposed pairing-free certificateless two-party authenticated key agreement protocols for Smart grid. According to our analysis, these two protocols are insecure against basic impersonation attacks of malicious key-generator centers, man-in-the-middle attacks of malicious key generator centers, and key offset attacks. We also found and pointed out some errors in the descriptions of these protocols.

## I. INTRODUCTION

While the Internet brings convenience to people's lives, it also places a higher demand on data transmission and storage. A smart grid (SG) is a fully automated power transmission network that monitors and controls every communication participant and grid node to ensure a bidirectional flow of information and power between all nodes. Recently, security incidents have frequently occurred in SGs. For example, an unauthorized user may intercept or change it while transmitting a power meter notification message. Such changes or leaks in messages in an SG network can have fatal consequences for the households to which the messages are sent. In other words, these threats allow unauthorized participants to obtain confidential personal information. From the SG model perspective, these types of attacks can negatively affect customers, markets, service providers, and operations. To prevent such security threats in an SG network, there must be a key agreement (KA) protocol to authenticate the identity of the communication participant or node and share the session key.

The KA protocol based on public key infrastructure (PKI) is a good option that can be used for secure identity authentication. However, PKI has the drawback of being expensive for certificate authorities (CA) to manage certificates. To reduce the cost of managing the certificate, the identity basic key agreement (IDKA) protocol has been proposed, in which identity information is the public key of the user. However, the IDKA protocol raises a key escrow problem that requires full trust in the key generation center (KGC). To solve these problems, certificateless key agreement (CKA) and certificateless authenticated key agreement (CAKA) protocols have been proposed, and certificateless two-party authenticated key agreement (CTAKA) is the basis of these protocols. The CTAKA protocol has been studied in two directions: one that uses pairing and the other that does not. Because the CTAKA protocol based on pairing requires a large amount of computation for pairing, research is being conducted to construct a protocol that reduces or does not use the number of pairings.

In general, KA protocols can be classified into three categories according to the trusted organization such as KGC or CA. A protocol is said to have a trust level of 1 if a trusted organization can calculate the private keys of all communication participants registered in the system. As discussed above, the IDKA protocol has trust level 1 because it has a key escrow problem, in which the communication participant must fully trust the KGC. The protocol is said to have trust level 2 if the organization cannot compute the private keys of the communication participants, but can create false assurances, such as fake public keys, to impersonate the participants registered in the system. Although many CAKA protocols have been proposed thus far, including the protocol proposed by Al-Riyami-Peterson [1], most are insecure and have a trust level of 2. Trust level 3 means that the organization cannot calculate the private key of the communication participant, and it is released when the KGC generates a fake assurance. Trust level 3 means that the organization cannot calculate the communication participant's private key, and if the organization creates fake assurance, it will be exposed. The KA protocol, which is based on the traditional PKI model, had a trust level of 3. This is because the public key of the communication participant is explicitly authenticated by the CA and the existence of two valid public keys (including certificates) for one ID indicates that the CA has been deceived. Therefore, designing a pairing-free CTAKA lightweight protocol with trust level 3 is a good option for enabling secure communication in an SG network.

In 2021, Cui et al. proposed a new pairing-free CTAKA protocol for the Internet of Things (IoT), and proved the security of the protocol in the improved extended Canetti-Krawczyk (eCK) security model. [2] The proposed protocol was programmed and implemented to compare its performance with those of other protocols. Deng et al. proposed a new pairing-free CTAKA lightweight protocol for SGs that uses only four scalar multiplications and proved its security in the standard model. [3]

*A. CONTRIBUTION*

In this paper, we conducted a security analysis on pairing-free CTAKA protocols for SGs proposed by Cui and Deng et al. According to our analysis, these protocols are insecure against the basic impersonation attack (BIA) of malicious KGCs, the man-in-the-middle attack (MMA) of malicious KGCs, and the key offset attack (KOA). We also found and pointed out some errors in the description of these protocols.

*B. ORGANIZATION OF PAPER*

The paper is structured as follows.

Section 2 briefly introduces the results of previous research on CTAKA protocols and discusses some protocols proposed for SGs.

In Section 3, some security properties to be satisfied by recently proposed AKA protocols, such as KOA resilience, BIA resilience, and MMA resilience, are mentioned.

Section 4 analyzes the security of the CTAKA protocol for power IoT proposed by Cui *et al.* To do this, we first briefly explain the protocol proposed by Cui *et al.* and then prove that this protocol cannot withstand MMA, BIA of malicious KGC, and KOA. We also found and corrected some notation errors in the protocol of Cui *et al.* Section 4 presents an analysis of the security of the SG protocol proposed by Deng et al.

Similarly, Deng *et al.*'s protocol was first described; then, this protocol was insecure against KOA and some errors in the protocol description were explained.

Section 5 presents the conclusions and directions for future research.

## II. RELATED WORKS

In 1976, the KA protocol proposed by Diffie and Hellman was vulnerable to MMA, because the identity of the communication participant could not be authenticated. [4] Law *et al.* proposed the KA protocol, which uses a digital certificate issued by a trusted third-party authentication authority and implicit key authentication using digitally signed tokens to verify public keys. [5] Therefore, this protocol employs an expensive PKI model for user authentication. In 1984, Shamir proposed IDKA to avoid certificate management in the previous PKI, and subsequently, some identity basic two-party key agreement (IDTAKA) protocols were proposed. [6] In IDKA, because the KGC holds the private keys of all communication participants, if there is a potentially malicious manager inside the KGC, the communication participant information is at risk. In 2003, Al-Riyami and Paterson published certificateless public key cryptography that solves the key escrow problem while avoiding certificate management, and the CKA protocol was first proposed. [1]

To use the KA protocol in actual applications, its security must be analyzed using a formal security model. The Bellare and Rogaway (BR) security model proposed in 1993, the modified BR (mBR) model proposed in 1995, and the BJM model do not have security attributes such as weak perfect forward security, key compromise impersonation (KCI) resilience, and ephemeral secret key leakage (ESKL) resilience. [7]–[9] In 2001, Canetti and Krawczyk (CK) proposed the CK model in which the adversary can query session state information for sessions other than the test session. [10] However, the CK model did not exhibit resilience to KCI or ESKL. To this end, Lamacchia *et al.* proposed an eCK model in 2007 that allows an adversary to query a clear combination of a static private key and real private key, even during an inspection session, thereby guaranteeing maximum recovery properties. [11] In 2010, Sarr *et al.* proposed an enhanced eCK (e2CK) model to capture intermediate computation leakages. [12] Many of the protocols proposed thus far have been proven to be secure in a formal security model.

Several CTAKA protocols based on pairing have been proposed. However, the computational cost of the pairing operation is approximately three times that of the scalar multiplication operation on an elliptic curve. In 2009, Geng and Zhang proposed the first pairing-free CTAKA protocol, many of which have been proposed since then. [13] In 2012, He *et al.* [14] proposed a pairing-free CTAKA scheme. However, Cheng [15] pointed out that this protocol is vulnerable to a type-I adversary that obtains the ephemeral private key of either party. Kim *et al.* [16] presented a pairing-free CTAKA protocol and argued that security can be demonstrated in the eCK model. However, in 2018, Bala *et al.* described this protocol as vulnerable to KCI attacks and proposed an improved protocol. [17] In 2019, Xie *et al.* proposed an efficient pairing-free CTAKA protocol under the gap Diffie Hellman (GDH) assumption. [18] However, in 2019, Renu *et al.* proved that the protocols of Bala *et al.* and Xie *et al.* are all vulnerable to the BIA of malicious KGCs and proposed a pairing-free CTAKA protocol with trust level 3. [19] Bala *et al.* proposed a pairing-free IDTAKA protocol based on the GDH assumption for a wireless network and proved its security in the eCK model. [20] However, in 2020, Renu *et al.* proved that the protocol proposed by Bala *et al.* was vulnerable to KOA and that protocols such as Islam do not guarantee partial forward secrecy (FS). [21], [22]

Several AKA protocols have been proposed and applied to SGs. In 2016, Tsai *et al.* proposed a new authentication protocol applicable to SGs [23], but in 2018, Mahmood *et al.* indicated that this protocol used computationally expensive bilinear pairing and proposed an improvement protocol. [24] In 2018, Li *et al.* proposed an improved SM2-based KA and mutual identification authentication protocol for SGs. [25] However, because these protocols are all implemented using the PKI model, certificate management is complicated and unsuitable for power IoT with a large number of terminals. Subsequently, Lin *et al.* proposed an improved secure communication protocol that strengthens the security of network communication by adding a timestamp and a digital signature to a message. [26] In 2020, Gupta *et al.* proposed a provably secure and lightweight IDTAKA protocol for industrial IoT environments. However, in 2021, Li *et al.* proved that this protocol is insecure against ephemeral KCI attacks. [27], [28]

## III. REQUIRED SECURITY PROPERTIES

This section describes the security properties of the CTAKA protocol considered in this study. [19], [22]

*Basic Impersonation Resilience* (BIR): This security property implies that an adversary cannot impersonate itself as a legitimate communication participant without knowing their static private key.

*Key Compromise Impersonation Resilience* (KCIR): Communication participant $A$ is said to be compromised if an attacker knows participant $A$'s static private key. This security property means that the adversary who has corrupted participant $A$ can easily impersonate itself as $A$, but cannot impersonate itself to participant $A$ as another uncorrupted participant $B$.

*Man-in-the-Middle Attack Resilience* (MMAR): The adversary intercepts the message sent by communication participants $A$ to $B$ and sends it after changing it. The adversary can then pretend to $A$ and share a key with $B$. Similarly, the adversary can share the key with $A$ by forging a message $B$ sent to $A$. However, at this time, communication participants $A$ and $B$ do not believe that they share the secret key with their adversary. Finally, the adversary shares a key with either $A$ or $B$, impersonating it as an honest participant. This security property implies that such a case should not be exist.

*Key Offset Attack Resilience* (KOAR): In a key offset attack, the adversary modifies the message sent by multiplying the ephemeral public key by a random value. Consequently, the final session key calculated by the communicating parties will be different. Many AKA protocols that do not verify the final session key are vulnerable to this attack. The difference from a man-in-the-middle attack is that an adversary does not use any s communication security information.

*Known-Key Security* (KKS): This security property means that different session keys are generated each time a protocol is executed. Therefore, even if an adversary knows the previous session key, he/she should not be able to retrieve the current session key.

*No Key Control* (NKC): This security property implies that a communication participant cannot pre-determine part or all session keys before the protocol is executed.

*Unknown Key-Share Resilience* (UKSR): This security property means that a communication participant who shares a session key with an honest participant cannot be forced into thinking that this key has been shared with an adversary. In other words, participant $A$ cannot be coerced into sharing a key with participant $B$ without $A$'s knowledge; that is, when $A$ believes that the key is shared with some participant $C \neq B$, and $B$ (correctly) believes the key is shared with $A$.

*Perfect Forward Secrecy* (PFS): This security property means that the past session key is secure even if all communication participants are compromised; that is, even if the adversary knows all the static private keys of the communication participants, it cannot calculate the previously established session key. KGC's perfect forward secrecy (KGC-PFS) means that past session keys remain secure, even if KGC's master secret key is leaked.

*Partial Forward Secrecy* (partial FS): This security property implies that if the adversary knows only the static private key of one communication participant, it cannot compute a previously established session key.

*Weak Perfect Forward Secrecy* (WPFS): This security property means that even if the participants' static private keys are compromised, the secrecy of previously established session keys is guaranteed, but only for sessions in which the adversary does not actively interfere.

*Ephemeral Private Key Leakage Resilience* (EPKLR): This security property means that even if the adversary knows both ephemeral private keys of the communication participants, it cannot compute the session key without further knowledge of either participant's static private key.

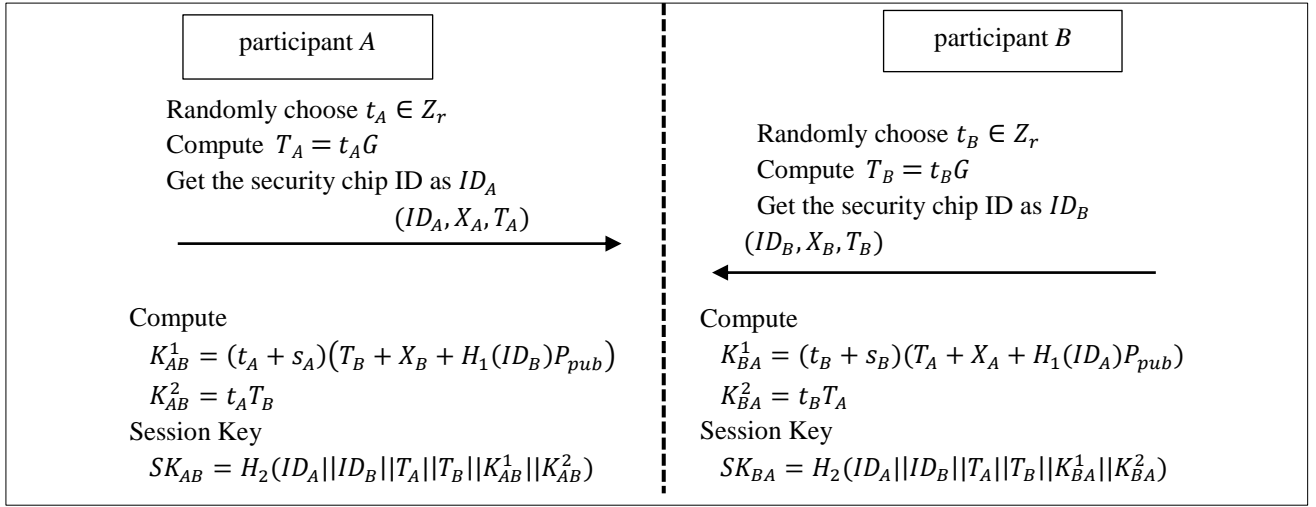

**FIGURE 1.** Authentication and Key Agreement process of the protocol proposed by Cui *et al.*

## IV. SECURITY ANALYSIS OF TWO RECENT PROTOCOLS FOR SGs

In this section, we analyze the security of two recent pairing-free CTAKA protocols for SGs proposed in [2], [3].

### A. SECURITY ANALYSIS OF CUI et al.'s PROTOCOL

Cui *et al.* proposed a new pairing-free CTAKA protocol that can support two-party authenticated key agreements between the power terminals and management systems. They argued that the e2CK security model and the proposed protocol under the assumption of Computation Diffie-Hellman (CDH) difficulty can achieve security. According to our analysis, the protocol proposed by Cui *et al.* is vulnerable to KOA, BIA by malicious KGC, and MMA by malicious KGC. First, we briefly review the protocol and then proceed with security analysis.

#### 1) A BRIEF REVIEW OF CUI et al.'s PROTOCOL

The protocol consists of five steps: initialization, private key generation, public key generation, and key agreement. A detailed description of the protocol below:

(1) Initialization

This step is to generate some public parameters for the protocol.
① KGC chooses one elliptic curve $E$.
② Next, KGC selects a random value $s \in Z_r$ as the master secret key.
③ KGC generates a master public key $Ppub = sG$ using $s$.
④ Two hash functions $H_1$ and $H_2$ are chosen for the public parameters, where $H_1: \{0,1\}^* \to Z_{r*}^*$ can map the participants' identity to the elements in $Z_r$, and the hash function $H_2: \{0,1\} \to \{0,1\}^k$ is chosen to compute the session key.
⑤ The public parameter is $PP = \{GF(q), G, E, P_{pub}, H_1, H_2\}$, and the KGC exposes the $PP$ to all participants in the system.

(2) Partial Private Key Generation
① If participant $i$ sends its $ID_i$ to the KGC, the KGC computes the partial private key $d_i = sH_1(ID_i)$.
② Next, KGC returns the key to the participant through a secret channel.

(3) Private Key Generation
① The participant $i$ chooses one random value $x_i \in Z_r$.
② Next, participant $i$ computes the private key $s_i = x_i + d_i$ where the partial private key $d_i$ is from KGC.

(4) Public Key Generation

In this step, participant $i$ computes $X_i = x_i G$ as its public key.

(5) Key Agreement

In this step, participant $A$ with identity $ID_A$ and participant $B$ with identity $ID_B$ can establish a connection and calculate the same session key after completing the processes of authentication and key agreement. Fig. 1 shows the complete process of the protocol proposed by Cui *et al.*

#### 2) KEY OFFSET ATTACK

In 1997, Blake-Wilson proposed the issue of an AKA protocol with shared key confirmation and mentioned the key-offset attack. [9]

After that, many authenticated key agreement protocols with shared key confirmation that are secure for KOA have been proposed. However, the protocol by Cui et al. is vulnerable to KOA.

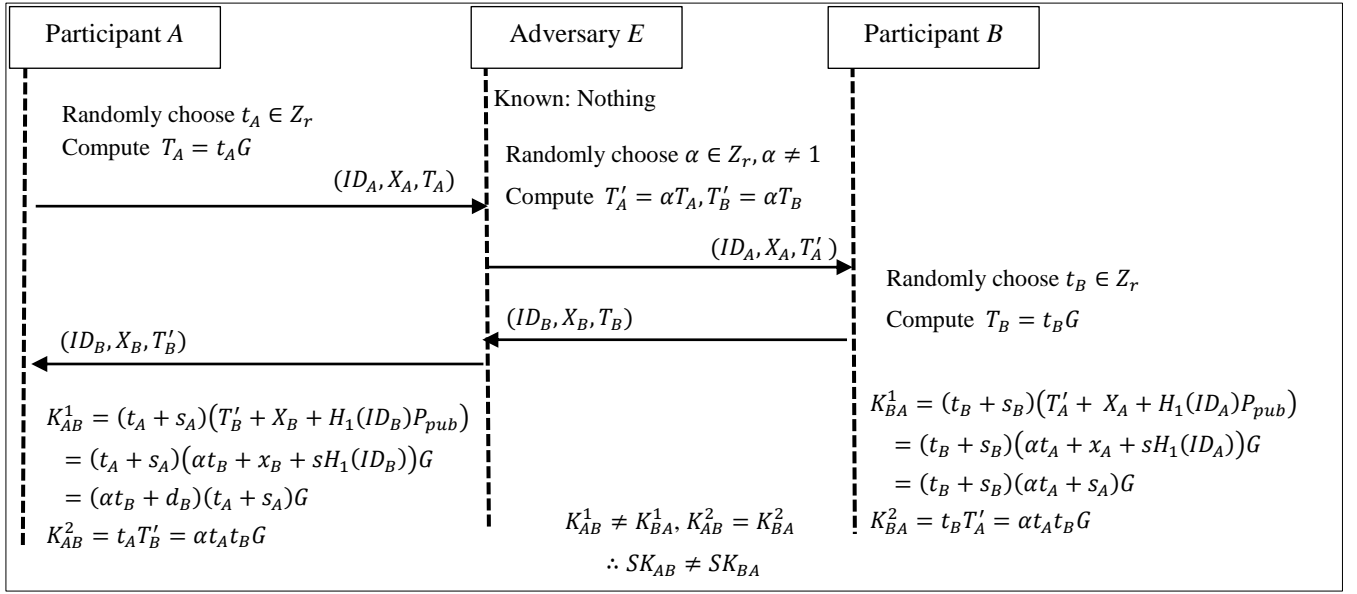

**FIGURE 2. Key Offset Attack against the protocol proposed by Cui *et al.*.**

The adversary first intercepts the pairs $(ID_A, X_A, T_A), (ID_B, X_B, T_B)$ exchanged between communication participants *A* and *B*. The adversary randomly chooses $\alpha \in Z_r, \alpha \neq 1$ and computes $T'_A = \alpha T_A, T'_B = \alpha T_B$. Next, the modified pair $(ID_A, X_A, T'_A), (ID_B, X_B, T'_B)$ is send to participants *B* and *A* respectively. Then $K^1_{AB} \neq K^1_{BA}, K^2_{AB} = K^2_{BA}$, so $SK_{AB} \neq SK_{BA}$. This means that the protocol lacks key integrity, and that the computed session key does not depend solely on the input provided by the protocol participant. From the attack, the adversary does not gain any knowledge of the agreed session key, but the two participants generate an incorrect session key.

The KOA is particularly effective in energy-constrained applications, such as smart meters and wireless body area networks. The adversary can repeat this attack and completely consume energy resources by preventing the communication participants from sharing the key. KOAR can be easily achieved by verifying the integrity of the shared key using a hash function. The KOA against the protocol described by Cui *et al.* is shown in Fig. 2.

#### 3) BASIC IMPERSONATION ATTACK BY MALICIOUS KGC

In this section, we prove that Cui *et al.*'s protocol is insecure against the BIA of a malicious KGC. A malicious KGC administrator who possesses a master key *s* can proceed with basic impersonation attacks as follows:

Because the malicious KGC does not know the private secret value $x_A$ of communication participant *A* registered in the system, it cannot know *A*'s static private key $s_A = x_A + d_A$. However, a malicious KGC knows *A*'s partial private key $d_A$. Therefore, malicious KGC randomly chooses $x'_A \in Z_r, t'_A \in Z_r$ and calculates $s'_A = x'_A + d_A, X'_A = x'_A G, T'_A = t'_A G$. Next, the malicious KGC sends $(ID_A, X'_A, T'_A)$ to B. In this way, a malicious KGC can impersonate itself as *A* without knowing *A*'s private key $s_A$. In other words, Cui *et al.*'s protocol is insecure against the basic impersonation attack of malicious KGCs. Note that malicious KGC cannot replace *A*'s public key, so it does not replace the $(ID_A, X_A, T_A)$ sent by *A*, but rather $(ID_A, X'_A, T'_A)$ from the beginning. The BIA against the protocol described by Cui *et al.* is shown in Fig. 3.

1) MAN-IN-THE-MIDDLE ATTACK BY MALICIOUS KGC

If the BIA discussed above is executed between two communication participants *A* and *B*, respectively, the malicious KGC administrator *E* can share the session key with *A* and *B*.

To launch the MMA, the malicious KGC can only intercept the messages of participants *A* and *B* on both sides, sending the modified pairs $(ID_A, X'_A, T'_A)$, $(ID_A, X'_B, T'_B)$. The malicious KGC can establish two connections, one with each drone. Fig. 4 shows the MMA by malicious KGC.

#### 4) INCORRECTION OF NOTATION IN cui *et al.*'s PROTOCOL

There were some errors in the protocol proposed by Cui et al.

First, in subsection 4.2 of [2], the private key of the private key generation step should be $s_i = x_i + d_i$, not $s_i = (x_i, d_i)$. This is because only the following expressions were established:

$$\begin{aligned}
K^1_{AB} &= (t_A + s_A)(T_B + X_B + H_1(ID_B)P_{pub}) \\
&= (t_A + x_A + d_A)(t_B + x_B + H_1(ID_B)s)G \\
&= (t_A + x_A + sH_1(ID_A))(t_B + x_B + d_B)G \\
&= (t_B + x_B + d_B)(t_A + x_A + H_1(ID_A)s)G \\
&= (t_B + s_B)(T_A + X_A + H_1(ID_A)P_{pub}) = K^1_{BA} \quad (1)
\end{aligned}$$

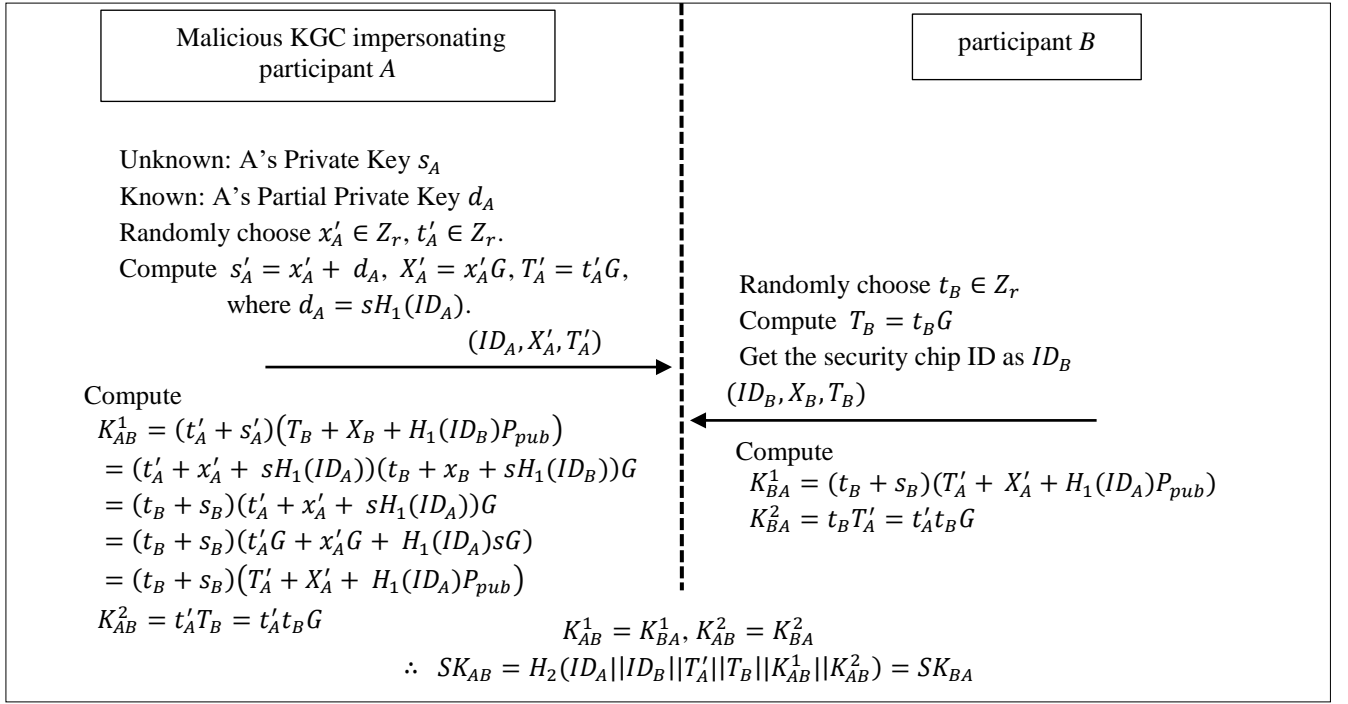

**FIGURE 3.** Basic Impersonation Attack against the protocol proposed by Cui *et al.*

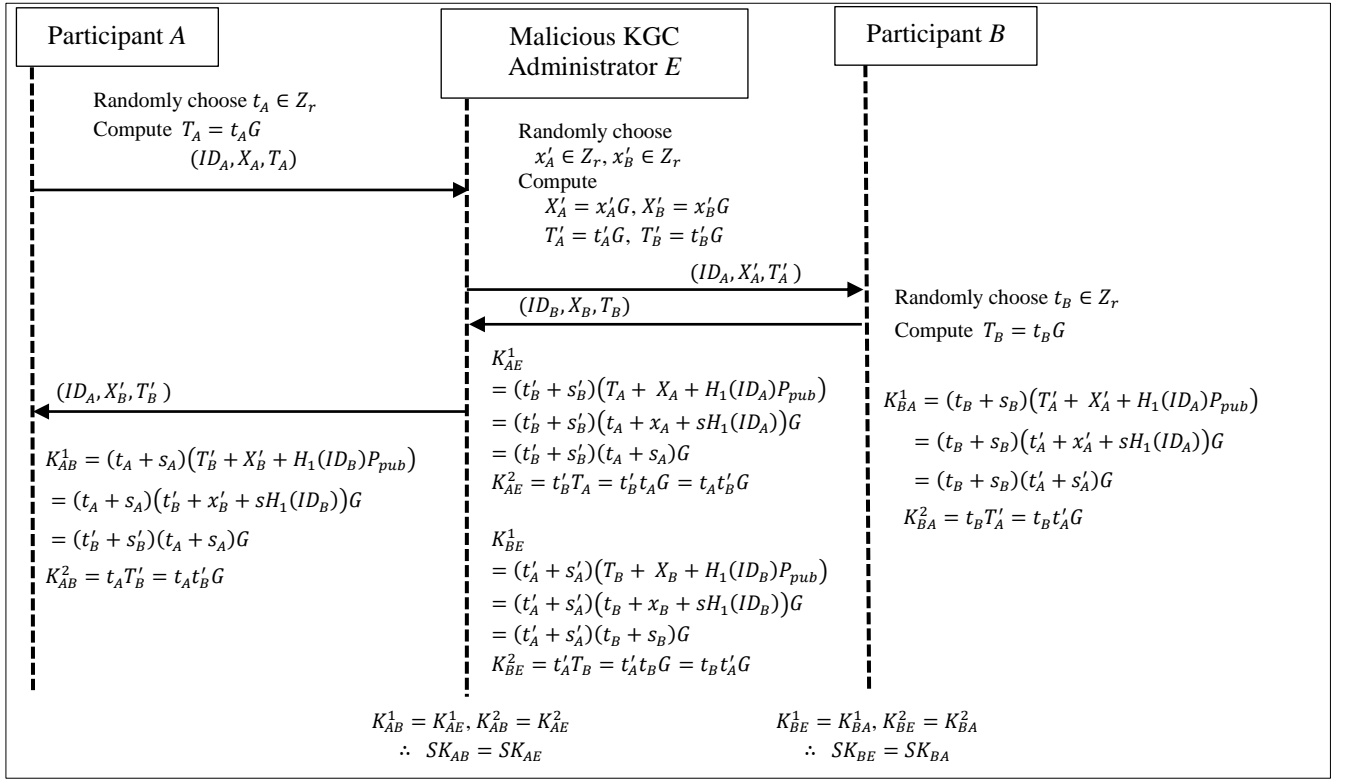

**FIGURE 4.** Man-in-the-Middle Attack by malicious KGC against the protocol proposed by Cui *et al.*

Next, in Fig. 1, it is $K_{BA}^1 = (t_B + s_B)(T_A + X_A + H_1(ID_A)P_{pub})$, not $K_{BA}^1 = (t_B + s_B)(T_B + X_B + H_1(ID_A)P_{pub})$.

### B. SECURITY ANALYSIS OF DENG et al.'s PROTOCOL

In [3], Deng *et al.* proposed a new CTAKA protocol for SG and proved the security of the CDH assumption and the eCK model. Security has been demonstrated in the standard model, but not in the random oracle model, and they say that the newly proposed protocol is more efficient than the previous schemes, as it does not require pairing and only requires four scalar multiplications.

However, in this section, we demonstrate the protocol of Deng et al. does not withstand KOA.

1) **A BRIEF REVIEW OF DENG et al.'s PROTOCOL**

(1) Setup phase

With a security parameter $m$, KGC generates system parameters as below.

① KGC chooses an additive group $G$ with the prime order $q$ defined on a curve $E/F_p$, and selects a generator $P$ of $G$.

② KGC selects the master random secret key $x \in Z_q^*$, and computes the systems public key $P_{pub} = xP$.

③ KGC chooses three secure hash functions $H_1: \{0,1\}^* \times G \times G \to Z_q^*$, $H_2: \{0,1\}^* \times \{0,1\}^* \times G \times G \times G \times G \times G \times G \to Z_q^*$ and $H_3: \{0,1\}^* \times \{0,1\}^* \times G \times G \times G \times G \times G \times G \times G \to Z_q^*$.

④ Finally, KGC publishes the system parameters $params = \{G, q, P, P_{pub}, H_1, H_2, H_3\}$, and keeps x secret.

(2) Registration phase

All be smart meters or service providers need to be registered with KGC. For a participant with identity $ID_k \in \{0,1\}^*$, the partial private key and public key are generated as follow.

① The participant chooses a random secret value $t_k \in Z_q^*$, and calculates public key $T_k = t_k P$, next sends the tuple $(ID_k, T_k)$ to KGC.

② KGC chooses a random secret value $r_k \in Z_q^*$, and calculates signature $R_k = r_k P$, $h_k = H_2(ID_k, T_k, R_k)$ and $d_k = r_k + h_k x$.

③ KGC send the partial private key pair $(d_k, R_k)$ to participant by a secure channel.

(3) Key agreement phase

The authentication and key sharing steps are illustrated Fig 5.

2) **KEY OFFSET ATTACK**

This section shows that Deng et al.'s protocol is also insecure against the KOA. The adversary first intercepts the pairs $(R_i, M_i)$ and $(R_j, M_j)$ exchanged between the smart meter $SM_i$ and service provider $SP_j$. The adversary accidentally chooses $\alpha \in Z_r, \alpha \neq 1$ and then computes $M_i' = \alpha M_i$, $M_j' = \alpha M_j$. Next, adversary sends the modified pairs $(R_i, M_i')$ and $(R_j, M_j')$ to $SP_j$ and $SM_i$, respectively.

$$l_{ij} = H_2(ID_i, ID_j, T_i, T_j, R_i, R_j, M_i, M_j')$$
$$\neq H_2(ID_i, ID_j, T_i, T_j, R_i, R_j, M_i', M_j) = l_{ji} \quad (2)$$
$$K_{ij} = (l_{ij}a_i + t_i + d_i)(l_{ij}M_j' + T_j + R_j + h_j P_{pub})$$
$$\neq (l_{ij}\alpha a_i + t_i + d_i)(l_{ij}M_j + T_j + R_j + h_j P_{pub})$$
$$= K_{ji} \quad (3)$$

Therefore, $SK_{AB} \neq SK_{BA}$. The KOA against the protocol proposed by Deng et al. is shown in Fig 6.

3) **INCORRECTION OF NOTATION IN DENG et al.'s PROTOCOL**

The authors claimed that their scheme is the first ID-based AKA scheme to demonstrate security in the standard model. Therefore, the communication participant's *ID* is its public key, and communication participant *A* has generated a public key $T_i$, but does not send it to communication participant *B*. However, in the key agreement process, communication participant *B* must calculate the shared key using *A*'s public key $T_i$. This question is similarly raised for communication participant *B*. Therefore, the communication participant must transmit the public key. However, if the communication participant's public key is transmitted as it is, this protocol is vulnerable to KCI attacks.

## V. CONCLUSION

To successfully operate in a fully automated power transmission network, secure communication must be ensured. For this purpose, many session key agreement protocols for secure communication in the SG environment have been proposed and applied. However, according to our analysis, several recently proposed protocols for SGs are vulnerable to various attacks.

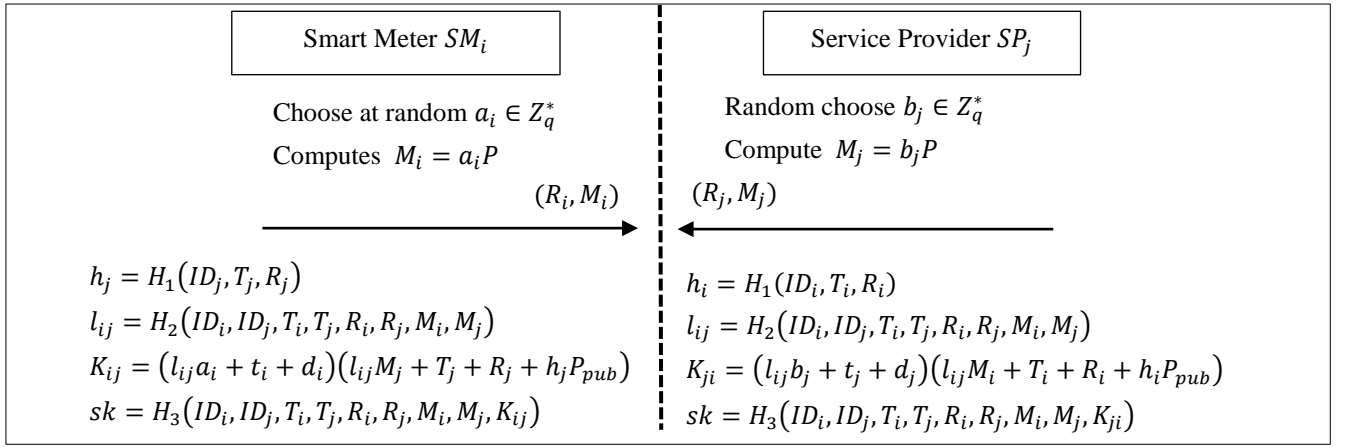

**FIGURE 5.** Authentication and Key Agreement process of the protocol proposed by Deng *et al*.

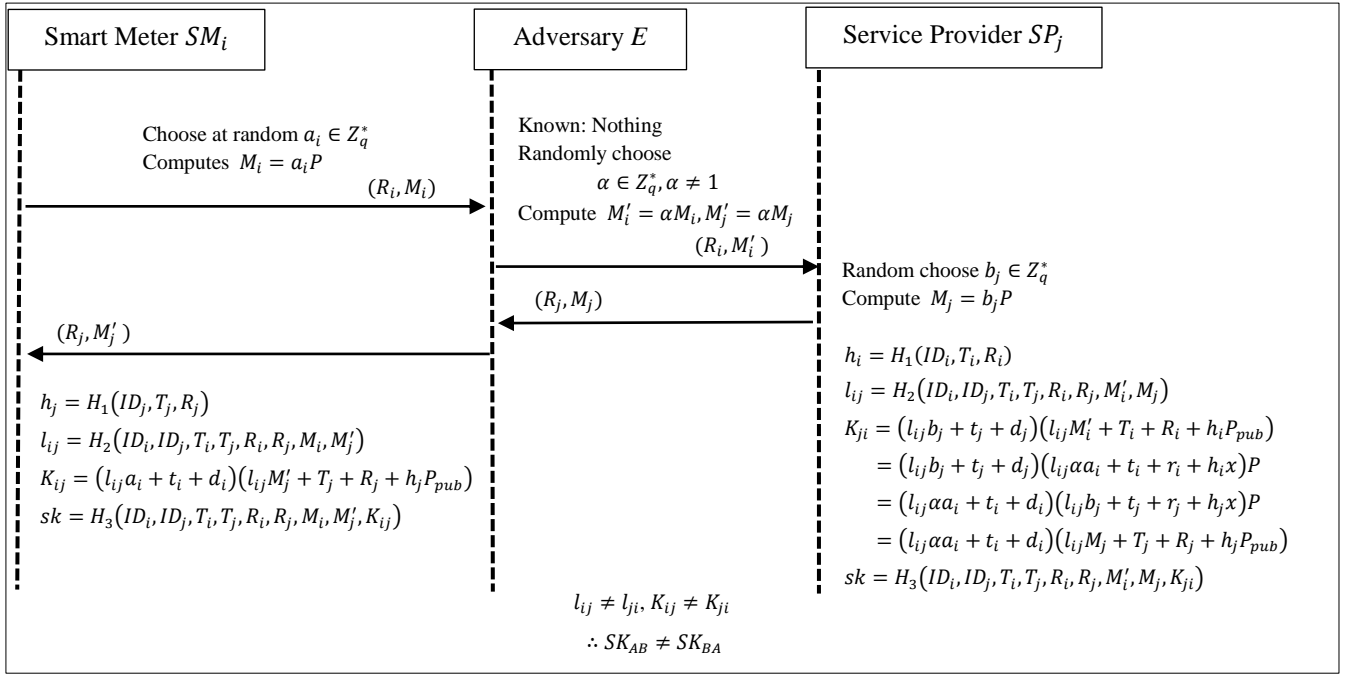

**FIGURE 6.** Key Offset Attack against the protocol proposed by Deng *et al.*

According to our analysis, Cui *et al.*'s protocol is trust level 2, which is insecure against malicious KGC's BIA and MMA, and both protocols are vulnerable to KOA. We also found and pointed out some errors in the descriptions of these protocols.

In the future, it will be desirable to design a lightweight protocol with trust level 3 that can withstand the attacks considered above while considering the characteristics of the SG.

**ACKNOWLEDGMENT**
We thank the anonymous reviewers, associate editors, and Editor-in-Chief for their valuable feedback on the paper.